\documentclass[prd,preprint,nofootinbib,showpacs,preprintnumbers,amsmath,amssymb]{revtex4}

\usepackage{bm}% bold math
\usepackage{hyperref}

\newcommand{\be}{\begin{equation}}
\newcommand{\ee}{\end{equation}}
\newcommand{\bea}{\begin{eqnarray}}
\newcommand{\eea}{\end{eqnarray}}
\newcommand{\p}{\partial}

\begin{document}

\title{Hawking radiation via anomaly cancelation for the black holes \\
 of five-dimensional minimal gauged supergravity}
\author{Achilleas P. Porfyriadis}
\email{apporfyr@mit.edu}
\affiliation{Department of Physics\\ Massachusetts Institute of Technology\\ Cambridge, Massachusetts 02139, USA}
\date{\today}

\begin{abstract}
The anomaly cancelation method proposed by Wilczek et al. is applied to the general charged rotating black holes in five-dimensional minimal gauged supergravity. Thus Hawking temperature and fluxes are found. The Hawking temperature obtained agrees with the surface gravity formula. The black holes have charge and two unequal angular momenta and these give rise to appropriate terms in the effective U(1) gauge field of the reduced (1+1)-dimensional theory. In particular, it is found that the terms in this U(1) gauge field correspond exactly to the correct electrostatic potential and the two angular velocities on the horizon of the black holes, and so the results for the Hawking fluxes derived here from the anomaly cancelation method are in complete agreement with the ones obtained from integrating the Planck distribution.
\end{abstract}


\maketitle

\section{Introduction}

Hawking radiation is a key effect in quantum gravity. There are various derivations, including Hawking's original one \cite{Hawking:1974sw} calculating Bogoliubov coefficients, which have shown that the radiation is thermal with a blackbody spectrum at a temperature given by the surface gravity, $T_H=\kappa/{2\pi}$. Derivations have been found also using Euclidean quantum gravity \cite{Gibbons:1976ue} and in string theory \cite{Strominger:1996sh}. A derivation using semiclassical methods (WKB) in a tunneling picture has been proposed in \cite{Parikh:1999mf}.

Recently, a new method for calculating Hawking radiation from the horizon of black holes has been developed by Robinson and Wilczek \cite{Robinson:2005pd} and by Iso, Umetsu, and Wilczek \cite{Iso:2006ut,Iso:2006wa}. The new method ties the existence of Hawking radiation to quantum anomalies at the horizon of black holes. More precisely, it is argued that Hawking fluxes are necessary in order to cancel gauge and gravitational anomalies at the horizon and restore invariance in the effective field theory under gauge and diffeomorphism transformations. The method uses a massless scalar field to probe the black hole background and the key step, the one that needs to be carried out separately for each black hole case, is the dimensional reduction of the scalar field action in the black hole background considered in each case. That is, one considers a complex scalar field in the black hole background and reduces its action near the horizon to a collection of $(1+1)-$dimensional fields. Then using the ideas and calculations in \cite{Robinson:2005pd} one obtains the Hawking temperature $T_H$ from the anomaly cancelation viewpoint. Furthermore, using the analysis carried out in \cite{Iso:2006ut,Iso:2006wa} one may also find the specific Hawking fluxes of the energy-momentum, angular momentum (if the black hole is rotating) and electric charge (if the black hole is charged).

During recent intense work the method proposed by Wilczek {\it et al.} \cite{Robinson:2005pd,Iso:2006wa,Iso:2006ut} has been successfully applied to various black objects \cite{various black holes}. In the direction of further extending the method, higher spin currents were studied under conformal transformations to obtain the full thermal spectrum of Hawking radiation \cite{higher spin}. An interesting point raised in \cite{Iso:2006wa} was that while it is the consistent form of the anomalies that is canceled, the boundary condition employed at the horizon is the vanishing of the covariant form of the current. Following up, in \cite{covariant anomaly "method"} a variant of the original method was proposed in which only covariant forms of the anomalies are used and all its applications \cite{covariant anomaly "apps"} have confirmed the results obtained in the corresponding papers in \cite{various black holes}. For a discussion of the relative merits of these various approaches see \cite{Umetsu:2008cm}. In this paper however, as we are primarily concerned with applying the method to a new black hole, we will follow the original approach of \cite{Robinson:2005pd,Iso:2006wa,Iso:2006ut} only. For other related work on the subject see also \cite{various}.

In this paper we apply the anomaly cancelation method \cite{Robinson:2005pd,Iso:2006wa,Iso:2006ut} to obtain the Hawking temperature and fluxes of the recently discovered general nonextremal charged rotating black holes in five-dimensional minimal gauged supergravity \cite{Chong:2005hr}. These black holes are the charged generalizations of the neutral Kerr-AdS black holes in $D=5$ gauged supergravity found in \cite{Hawking:1998kw}. They are parametrized by four nontrivial parameters, namely, their mass, charge, and two independent rotation parameters in the two orthogonal 2-planes. The special case in which the two rotation parameters are equal was obtained first in \cite{Cvetic:2004hs,Cvetic:2004ny}. $D=5$ gauged supergravities are especially interesting in the context of the AdS/CFT correspondence, as their properties are related to those of the strongly coupled conformal field theories on the four-dimensional boundary of AdS$_5$. In particular, it has been shown in \cite{Hawking:1998kw} that the backgrounds of rotating asymptotically AdS black hole solutions of $D=5$ gauged supergravities are dual to $D=4$ conformal field theories in the rotating Einstein universe on the boundary. It is therefore an interesting and important task to study the thermodynamic properties of the black holes in gauged supergravities. In this paper the anomaly cancelation method is applied for the first time to black holes in supergravity theory.

The rest of the paper is as follows. In Section \ref{Section: The Black Hole} we give the metric and basic properties of the charged rotating black holes in $D=5$ minimal gauged supergravity. We also derive the fluxes of Hawking radiation from general charged rotating black holes by integrating the Planck distribution. This will serve as a reference for comparing with our results obtained later using the anomaly cancelation method. In Section \ref{Section: anomaly cancelation} we use the dimensional reduction technique of \cite{Iso:2006ut} for the general $D=5$ minimal gauged supergravity black holes, and we apply the anomaly cancelation method, following \cite{Robinson:2005pd,Iso:2006ut,Iso:2006wa}, to derive the Hawking temperature and fluxes of charge, angular momenta, and energy-momentum tensor. We find agreement of the results with the surface gravity formula for the temperature and the results of section \ref{Section: The Black Hole} for the fluxes. Section \ref{Section: Conslusion} is the conclusion. In the Appendix we give some details of the calculations involved in the dimensional reduction of the $D=5$ minimal gauged supergravity black holes.

\section{Charged rotating black holes in $D=5$ minimal gauged supergravity}\label{Section: The Black Hole}

In this section we first review the space-time solution and basic properties of the black holes we are considering. We follow closely the notations in \cite{Chong:2005hr}. Here we also find the Hawking fluxes derived from integrating the Planck distribution.

Minimal gauged five-dimensional supergravity is described by the Lagrangian density
\be
{\cal L} = (R+ 12g^2)\, {*\rlap 1\mkern4mu{\rm l}} - \frac{1}{2} {*F}\wedge F + \frac{1}{3\sqrt3}F\wedge F\wedge A\,,
\ee
where $F=dA$, and the gauge-coupling constant $g$ is assumed to be positive, without loss of generality. Solutions to the corresponding equations of motion, which describe general nonextremal charged rotating black holes with two independent rotation parameters, were recently obtained in \cite{Chong:2005hr}. In Boyer-Lindquist type coordinates $x^\mu= (t, r, \theta,\varphi,\psi)$ that are asymptotically static\footnote{For the metric in Boyer-Lindquist coordinates which are rotating at spatial infinity see, for example, \cite{Aliev:2008yk}.}, the metric is given by
\bea\label{metric}
ds^2 &=& -\frac{\Delta_\theta\, [(1+g^2 r^2)\rho^2 d t + 2q \nu] \, d t}{\Xi_a\, \Xi_b \, \rho^2} + \frac{2q\, \nu\omega}{\rho^2} + \frac{\lambda}{\rho^4}\Big(\frac{\Delta_\theta \, d t}{\Xi_a\Xi_b} - \omega\Big)^2 + \frac{\rho^2 d r^2}{\Delta_r} + \frac{\rho^2 d\theta^2}{\Delta_\theta}\nonumber\\
&& + \frac{r^2+a^2}{\Xi_a}\sin^2\theta d\varphi^2 + \frac{r^2+b^2}{\Xi_b} \cos^2\theta d\psi^2\,,
\eea
where
\bea
\nu &=& b\sin^2\theta d\varphi + a\cos^2\theta d\psi\,, \qquad \omega =
a\sin^2\theta \frac{d\varphi}{\Xi_a} + b\cos^2\theta
\frac{d\psi}{\Xi_b}\,,\nonumber\\
\rho^2 &=& r^2 + a^2 \cos^2\theta + b^2 \sin^2\theta, \qquad
\Delta_\theta = 1 - a^2 g^2 \cos^2\theta - b^2 g^2
\sin^2\theta\,,\nonumber\\
\Delta_r &=& \frac{(r^2+a^2)(r^2+b^2)(1+g^2 r^2) + q^2 +2ab q}{r^2} - 2m
\,,\nonumber \\
\lambda&=& 2 m \rho^2 - q^2 + 2 a b q g^2 \rho^2, \qquad \Xi_a =1-a^2
g^2\,,\quad \Xi_b = 1-b^2 g^2\,.
\eea
The background gauge electromagnetic field is given by
\be\label{gauge potential A}
A = \frac{\sqrt3 q}{\rho^2}\, \Big(\frac{\Delta_\theta\, d t}{\Xi_a\, \Xi_b} - \omega\Big)\,.
\ee
Also note that the metric on the unit sphere is taken $d\Omega_3^2=d\theta^2+\sin^2\theta d\varphi^2+\cos^2\theta d\psi^2$, and so here $\theta$ runs over the range $0$ to $\pi/2$, and $\varphi, \psi$ take values between $0$ and $2\pi$.

In the appropriate parameter ranges, the metric \eqref{metric} describes regular rotating black holes, with no naked singularities or closed timelike curves \cite{Chong:2005hr}. The event horizon $r_+$ is the largest positive root of $\Delta_r=0$. The angular velocities at the horizon are given by
\be\label{horizon ang velocities}
\Omega_{a,h} = \frac{a(r_+^2+ b^2)(1+g^2 r_+^2) + b q}{(r_+^2+a^2)(r_+^2+b^2) + ab q}\,, \qquad
\Omega_{b,h} = \frac{b(r_+^2+ a^2)(1+g^2 r_+^2) + a q}{(r_+^2+a^2)(r_+^2+b^2) + ab q}\,.
\ee
There are three commuting Killing vectors, namely $\p_t,\p_\varphi$, and $\p_\psi$, associated with the translational and bi-azimuthal rotational isometries of \eqref{metric}. Thus using the corotating Killing vector
\be\label{Corotating Killing vector}
\ell = \p_t + \Omega_{a,h}\, \p_\varphi + \Omega_{b,h} \, \p_\psi\,,
\ee
we find that the electrostatic potential on the horizon is
\be\label{electrostatic potential on the horizon}
\Phi_h= \ell^\mu A_\mu\vert_{r_+}=\frac{\sqrt{3}qr^2}{(r_+^2+ a^2)(r_+^2+b^2)+abq}\,.
\ee

In order to compare with our result from the anomaly cancelation later, we give here the Hawking temperature as it was calculated in \cite{Chong:2005hr} via the surface gravity:
\be\label{temperature as surface gravity}
T_H=\frac{\kappa}{2\pi} = \frac{1}{2\pi}\frac{r_+^4[1+ g^2(2r_+^2 + a^2+b^2)] -(ab + q)^2}{r_+\, [(r_+^2+a^2)(r_+^2+b^2) + abq]}\,.
\ee
Let us also calculate here the Hawking fluxes one obtains from integrating the Planck distribution for a general rotating black hole with a nontrivial electromagnetic background gauge field $A$. The appropriate chemical potentials for fields radiated with azimuthal angular momenta $m$ and $n$ and with an electric charge $e$, are the horizon angular velocities $\Omega_{a,h}$ and $\Omega_{b,h}$ and the electrostatic potential on the horizon $\Phi_h$ (respectively). For fermions, the Planck distribution for blackbody radiation moving in the positive $r$ direction at a temperature $T_H$ is given by,
\be\label{Planck distribution}
N_{e, m, n}(\omega)= \frac{1}{e^{(\omega - e\Phi - m\Omega_{a,h}-n\Omega_{b,h})/T_H}+1}.
\ee
We consider only fermions in order to avoid superradiance \cite{Iso:2006ut}. From the above distribution and including the contribution from the antiparticles, we find the following Hawking fluxes of electric charge, angular momenta, and energy-momentum tensor respectively:
\begin{align}
F_q &= e \int_0^\infty \frac{d\omega}{2\pi} \left(N_{e,m,n}(\omega) - N_{-e, -m,-n}(\omega)\right)=\frac{e}{2\pi}\left(e\Phi_h + m\Omega_{a,h}+n\Omega_{b,h}\right), \label{flux el charge from Planck} \\
F_a &= m \int_0^\infty \frac{d\omega}{2\pi} \left(N_{e,m,n}(\omega) - N_{-e, -m, -n}(\omega)\right)=\frac{m}{2\pi}\left(e\Phi_h + m\Omega_{a,h}+n\Omega_{b,h}\right), \label{flux ang momentum a from Planck} \\
F_b &= n \int_0^\infty \frac{d\omega}{2\pi} \left(N_{e,m,n}(\omega) - N_{-e, -m, -n}(\omega)\right)=\frac{n}{2\pi}\left(e\Phi_h + m\Omega_{a,h}+n\Omega_{b,h}\right), \label{flux ang momentum b from Planck} \\
F_T &= \int_0^\infty \frac{d\omega}{2\pi}~\omega \left(N_{e,m,n}(\omega) + N_{-e, -m, -n}(\omega)\right)=\frac{1}{4\pi}\left(e\Phi_h + m\Omega_{a,h}+n\Omega_{b,h}\right)^2+ \frac{\pi}{12}T_H^2\,. \label{flux e-m tensor from Planck}
\end{align}
We shall derive the above fluxes by the anomaly cancelation method for the charged rotating black holes (\ref{metric}-\ref{gauge potential A}) in the following section.

\section{Dimensional reduction and anomaly cancelation} \label{Section: anomaly cancelation}

In this section we perform the dimensional reduction and anomaly cancelation procedure and derive the Hawking temperature and fluxes for the charged rotating black holes of $D=5$ minimal gauged supergravity.

Consider a massless complex scalar field in the background of (\ref{metric}-\ref{gauge potential A}). The free part of the action is
\bea
S&=&-\int d^5x\sqrt{-g}\,g^{\mu\nu}(D_\mu\phi)^*\,D_\nu\phi=\int d^5x \,\phi^* D_\mu(\sqrt{-g}\,g^{\mu\nu}D_\nu\phi) \nonumber \\
&=&\int d^5x\sqrt{-g}\, \phi^* \Big(g^{t t}D_t^2+2g^{t \varphi}D_tD_\varphi+2g^{t \psi}D_tD_\psi+g^{\varphi \varphi}D_\varphi^2+
g^{\psi \psi}D_\psi^2+2g^{\varphi \psi}D_\varphi D_\psi  \nonumber \\
&&\qquad\qquad\qquad\qquad +\frac{1}{\sqrt{-g}}\p_\theta \sqrt{-g}g^{\theta \theta}\p_\theta+\frac{1}{\sqrt{-g}}\p_r \sqrt{-g}g^{r r}\p_r \Big)\, \phi \,, \label{free action with g's}
\eea
where $D_t=\p_t+ieA_t$, $D_\varphi=\p_\varphi+ieA_\varphi$ and $D_\psi=\p_\psi+ieA_\psi$, and $e$ is the electric charge of $\phi$.
To study the near horizon theory, $r \to r_+$, it is most convenient to transform first to a ``tortoise'' like coordinate.
In our case if we first transform to the $r_*$ coordinate defined by
\be \label{tortoise}
dr=\frac{r^2 \Delta_r}{(r^2+a^2)(r^2+b^2)+abq} dr_*\equiv f(r)dr_*\,,
\ee
and then take the near horizon limit $r \to r_+$ (\emph{i.e.} $\Delta_r\to 0$) we obtain a near horizon action which, quite remarkably, may be written as\footnote{Refer to the Appendix for some of the calculations that lead to the given result.}:
\be\label{near horizon S with old theta}
S\approx -\frac{1}{2\Xi_a\Xi_b}\int dt dr_* \sin2\theta d\theta d\varphi d\psi \, \Psi \, \phi^* \left[(D_t+\Omega_a D_\varphi+\Omega_b D_\psi)^2-\p_{r_*}^2\right]\phi\,,
\ee
with
\be\label{ang velocities}
\Omega_{a} = \frac{a(r^2+ b^2)(1+g^2 r^2) + b q}{(r^2+a^2)(r^2+b^2) + ab q}\,, \qquad
\Omega_{b} = \frac{b(r^2+ a^2)(1+g^2 r^2) + a q}{(r^2+a^2)(r^2+b^2) + ab q}\,,
\ee
and
\be
\Psi=\frac{(r^2+a^2)(r^2+b^2)+abq}{r}\,.
\ee
Note that $D_t+\Omega_a D_\varphi+\Omega_b D_\psi=\p_t+\Omega_a \p_\varphi+\Omega_b \p_\psi+ie\Phi\,,$ with
\be
\Phi=\frac{\sqrt{3}qr^2}{(r^2+ a^2)(r^2+b^2)+abq}\,.
\ee
At this point we also wish to change the variable $\theta$ to $2\theta$ so that it runs over the range $0$ to $\pi$. Then \eqref{near horizon S with old theta} becomes,
\be
S\approx -\frac{1}{4\Xi_a\Xi_b}\int dt dr_* \sin\theta d\theta d\varphi d\psi \, \Psi \, \phi^* \left[(\p_t+\Omega_a \p_\varphi+\Omega_b \p_\psi+ie\Phi)^2-\p_{r_*}^2\right]\phi\,.
\ee
Here $\theta$ runs over the range $0$ to $\pi$, and $\phi, \psi$ take values between $0$ and $2\pi$ as before.
Then by expanding the complex field $\phi$ as
\be\label{expansion}
\phi=\sum_{l,m,n} \phi_{lmn}(r_*,t)\, P_{lm}(\cos\theta)e^{i m \varphi}e^{i n \psi}\,,
\ee
we may integrate out all dependence on angles to find that
\be
S\approx \sum_{l,m,n} -\frac{1}{4\Xi_a\Xi_b}\int dt dr_*  \, \Psi \, \phi_{lmn}^* \left[(\p_t+im\Omega_a +in\Omega_b+ie\Phi)^2-\p_{r_*}^2\right]\phi_{lmn}\,.
\ee
This is the most important equation of the paper. It implies that the near horizon theory has been reduced to an infinite collection of massless $(1+1)-$dimensional
fields labeled by the quantum numbers $l,m,n$. Indeed, transforming back to the original $(r,t)$ coordinates, to each of the fields $\phi_{lmn}$ corresponds the action
\be
S_{lmn}=\int dt\, dr\, \, \tilde{\Psi} \, \phi_{lmn}^* \left[-\frac{1}{f}(\p_t+im\Omega_a +in\Omega_b+ie\Phi)^2+
\p_r f \p_r\right]\phi_{lmn}\,,
\ee
where $f(r)={dr}/{dr_*}$ defined in \eqref{tortoise}, and $\tilde{\Psi}={\Psi}/({4\Xi_a\Xi_b})$. That is, each $\phi_{lmn}$ can be considered as a $(1+1)-$dimensional complex scalar field in the backgrounds of the dilaton $\tilde{\Psi}$, metric
\be\label{reduced metric}
ds^2=-f(r) dt^2+\frac{1}{f(r)}dr^2\, ,
\ee
and $U(1)$ gauge potential
\be\label{effective A}
{\cal A}_t=-e\Phi-m\Omega_a-n\Omega_b\,, \qquad {\cal A}_r=0\,.
\ee
In this effective $U(1)$ gauge field the first term is due to the electric field of the black hole and the two remaining ones are induced from the bi-azimuthal symmetry of the metric \eqref{metric}.

Also note that any possible mass or interaction terms in the full action would have been trivially suppressed in the near horizon limit too, since they would all be multiplied by a factor $f$ in the tortoise coordinates.

At this point we impose the constraint that the classically irrelevant ingoing modes vanish near the horizon and thus the theory becomes chiral. Hence anomalies arise, and for their cancelation one needs fluxes which, as will be shown, may account for the Hawking fluxes of the black hole. In particular, it was shown in \cite{Robinson:2005pd} that for a metric of the form \eqref{reduced metric} the cancelation, in the effective field theory, of the (purely timelike) anomaly for the energy-momentum tensor,
\be\label{E-M anomaly}
\nabla_\mu T^\mu_\nu =
 \frac{1}{96\pi\sqrt{-g}}\epsilon^{\beta\delta}
 \partial_\delta\partial_\alpha\Gamma^\alpha_{\nu\beta}
\ee
at the horizon, requires a total flux equal to that of a (1+1) dimensional beam of massless blackbody radiation moving in the positive $r$ direction at the temperature $T_H=\frac{1}{4\pi}\p_r f\vert_{r_+}$. It is worth noting that one can only obtain the total flux and not the precise spectrum so that blackbody radiation is an assumption rather than a derivation (see however \cite{higher spin}) The corresponding temperature $T_H$ has been checked to agree with the Hawking temperature for every space-time considered so far. Elevating this consistency check to a derivation we thus find that the Hawking temperature of our black holes is given by
\bea
T_H&=&\frac{1}{4\pi}\p_r f\vert_{r_+}=\frac{1}{4\pi}\frac{r^2}{(r^2+a^2)(r^2+b^2)+abq}\,\p_r \Delta_r\vert_{r_+} \nonumber \\ &=&\frac{1}{2\pi}\frac{r_+^4[1+ g^2(2r_+^2 + a^2+b^2)] -(ab + q)^2}{r_+\, [(r_+^2+a^2)(r_+^2+b^2) + abq]}\,.
\eea
This result agrees with \eqref{temperature as surface gravity} as given in \cite{Chong:2005hr} where the Hawking temperature was calculated as the surface gravity at the horizon $\kappa$ divided by $2\pi$.

In \cite{Iso:2006ut,Iso:2006wa} fluxes of  electric charge, angular momentum, and energy momentum-tensor
necessary for canceling, in the effective field theory, gauge and gravitational anomalies at the horizon were computed by deriving and solving the appropriate Ward identities. For example, the effective two-dimensional current $j^r$ associated with the original gauge symmetry, which is defined from the five-dimensional current $J^r$ by integrating over a 3-sphere
\be\label{2d current of original gauge symmetry}
j^r=\int d\Omega_3 \left(\frac{r\rho^2}{\Xi_a\Xi_b}\right)J^r\,,
\ee
is anomalous near the horizon, the gauge anomaly being given by
\be\label{anomalous eqn for 2d current of original gauge symmetry}
\p_r j^r=\frac{e}{4\pi}\p_r {\cal A}_t\,.
\ee
Note that the current \eqref{2d current of original gauge symmetry} is the consistent current, and the right-hand side of \eqref{anomalous eqn for 2d current of original gauge symmetry} is the gauge anomaly in a consistent form.
By solving this equation with appropriate boundary conditions (vanishing of the covariant form of current at the horizon) \cite{Iso:2006ut,Iso:2006wa}, the flux of electric charge from the black holes is obtained:
\be\label{flux el charge from anomalies}
-\frac{e}{2\pi}{\cal A}_t(r_+)=\frac{e}{2\pi}\left(e\Phi_h+m\Omega_{a,h}+n\Omega_{b,h}\right)\,.
\ee
Similarly, the currents $J^r_\varphi, J^r_\psi$ in the two-dimensional theory associated with the bi-azimuthal symmetries\footnote{These are defined in terms of the $T^r_\varphi, T^r_\psi$ components of the five-dimensional energy-momentum tensor as follows:
\[
J^r_\varphi=-\int d\Omega_3 \left(\frac{r\rho^2}{\Xi_a\Xi_b}\right)T^r_\varphi\,, \qquad
J^r_\psi=-\int d\Omega_3 \left(\frac{r\rho^2}{\Xi_a\Xi_b}\right)T^r_\psi
\]} obey analogous anomalous equations near the horizon,
\be
\p_r J^r_\varphi=\frac{m}{4\pi}\p_r {\cal A}_t\,, \qquad \p_r J^r_\psi=\frac{n}{4\pi}\p_r {\cal A}_t\,,
\ee
which lead \cite{Iso:2006ut,Iso:2006wa} to the following fluxes of angular momenta:
\bea
-\frac{m}{2\pi}{\cal A}_t(r_+)&=&\frac{m}{2\pi}\left(e\Phi_h+m\Omega_{a,h}+n\Omega_{b,h}\right)\,, \label{flux ang momentum a from anomalies}\\
-\frac{n}{2\pi}{\cal A}_t(r_+)&=&\frac{n}{2\pi}\left(e\Phi_h+m\Omega_{a,h}+n\Omega_{b,h}\right)\,. \label{flux ang momentum b from anomalies}
\eea
Finally, the anomalous equation of the two-dimensional energy-momentum tensor near the horizon is given by
\be
\p_r T^r_t={\cal F}_{rt}{\cal J}^r+{\cal A}_t\p_r {\cal J}^r+\p_r N^r_t\,,
\ee
where ${\cal F}_{rt} = \partial_r {\cal A}_t$ and ${\cal J}^r \equiv \frac{1}{e}j^r=\frac{1}{m}J^r_\varphi=\frac{1}{n}J^r_\psi$ satisfies $\p_r {\cal J}^r = \frac{1}{4\pi}\p_r {\cal A}_t$. Solving it determines the flux of energy-momentum, which is found to be $\frac{1}{4\pi}{\cal A}^2_t(r_+)+N^r_t(r_+)$ with $N^r_t=({f^{'}}^2+f\,f^{''})/192\pi$ \cite{Iso:2006ut,Iso:2006wa}. That is, in our case we find that the flux of energy-momentum is
\be\label{flux e-m tensor from anomalies}
\frac{1}{4\pi}{\cal A}^2_t(r_+)+N^r_t(r_+)=\frac{1}{4\pi}\left(e\Phi_h+m\Omega_{a,h}+n\Omega_{b,h}\right)^2+\frac{\pi}{12}T_H^2\,.
\ee
As (\ref{flux el charge from anomalies}, \ref{flux ang momentum a from anomalies}, \ref{flux ang momentum b from anomalies}, \ref{flux e-m tensor from anomalies}) clearly agree with the corresponding ones (\ref{flux el charge from Planck}, \ref{flux ang momentum a from Planck}, \ref{flux ang momentum b from Planck}, \ref{flux e-m tensor from Planck}) we have shown that the Hawking fluxes of the black holes considered in this paper may be derived from anomaly cancelation at the horizon of the black holes, and the results agree with those calculated from integrating the thermal spectrum \eqref{Planck distribution}.

\section{Conclusion}\label{Section: Conslusion}

In this paper we have applied the method of canceling quantum anomalies at the horizon to derive the Hawking temperature and fluxes of the general nonextremal charged rotating black holes with two independent angular momenta in $D=5$ minimal gauged supergravity. It was shown that near the horizon the quantum field behaves as an infinite set of two-dimensional conformal fields labeled by three quantum numbers. The effective two-dimensional theory near the horizon is described by charged matter fields in an electric field. In particular, the bi-azimuthal symmetry of the black holes leads to $U(1)$ gauge symmetries for each partial field mode, and their respective $U(1)$ charges are given by the corresponding azimuthal quantum numbers. By demanding gauge and diffeomorphism invariance in the effective two-dimensional field theory near the horizon, \emph{i.e.} by canceling the gauge and gravitational anomalies that arise upon suppressing classically irrelevant ingoing modes, we have calculated the Hawking temperature and fluxes of the black holes. The results found are consistent with the surface gravity formula and the fluxes obtained from integrating the Planck distribution, respectively.

The method of Wilczek \emph{et al.} adopted here uses only quantum anomalies at the horizon, and therefore it is quite universal, in the sense that it does not depend on the details of the quantum fields away from the horizon. Though not well understood why, it seems it is always possible to reduce a scalar field action in the background space-time of a black hole to an infinite sum of two-dimensional conformal field actions near the horizon. Once the dimensional reduction is accomplished, Hawking radiation may be calculated, and so far agreement with the Planckian results has been found in all cases where the method has been applied, including this paper. An important further development of the method would be calculating also the other universal thermodynamic quantity of black holes, namely, their entropy. Two-dimensional near horizon conformal field theory has been used to calculate the entropy of black holes via the Cardy formula quite successfully in the past \cite{Strominger:1997eq,Carlip:1998wz,Solodukhin:1998tc}. We might therefore hope that similar methods may lead to calculation of black hole entropy in the framework of the present method too.

\acknowledgments
I am grateful to Frank Wilczek and Sean Robinson for useful discussions on the subject and for reading this manuscript.

\appendix*
\section{Near horizon limit of the action}\label{Appendix}
In this appendix we give some additional details on the calculations involved in the passage from the action \eqref{free action with g's} to its near horizon limit \eqref{near horizon S with old theta}.

For the determinant of the metric \eqref{metric} we find
\be
\sqrt{-g} = \frac{r \rho^2 \sin 2\theta}{2 \Xi_a \Xi_b}\,.
\ee
Note that this is independent of the charge parameter $q$, and therefore it is the same for both charged and uncharged black holes. The inverse of the metric \eqref{metric} may be written as \cite{Davis:2005ys}
\begin{align}
\rho^2 \, g^{t t} = {}& -\frac{ (a^2 + b^2) (2 m r^2 - q^2)}{r^2 \Delta_r}
  -\frac{ (r^2+ a^2) (r^2 + b^2) [ r^2(1 - g^2 (a^2 + b^2)) - a^2
  b^2 g^2] }{ r^2 \Delta_r } \nonumber
\\ & - \frac{2 m a^2 b^2}{r^2 \Delta_r }- \frac{ 2 a b q r^2 }{ r^2
  \Delta_r } - \frac{ a^2 \cos^2 \theta \, \Xi_a + b^2 \sin^2 \theta
  \, \Xi_b}{ \Delta_{\theta} }\,, \nonumber  \\
\rho^2 \, g^{t \varphi} = {}& \frac{ a q^2 - [2 m a + b q (1 + a^2 g^2)]
 (r^2 + b^2)}{ r^2 \Delta_r }\,, &\nonumber \\
\rho^2 \, g^{t \psi} = {}& \frac{ b q^2 - [2 m b + a q ( 1 + b^2 g^2)]
  (r^2+ a^2)}{ r^2 \Delta_r }\,,  \nonumber \\
\rho^2 \, g^{\varphi \varphi} = {}&  \frac{ a^2 g^2 q^2}{ r^2 \Delta_r } +
\frac{\Xi_a}{\sin^2 \theta} + \frac{\Xi_a}{ r^2 \Delta_r } (1 + g^2
r^2) (r^2+ b^2) (b^2- a^2)   \nonumber \\
& - \frac{2 m}{ r^2 \Delta_r } (a^2 g^2 r^2
+ b^2)- \frac{2 a b q}{ \Xi_b \, r^2 \Delta_r } (\Xi_b \, g^2 (r^2 - a^2) -
  b^4 g^4 +1)\,, \nonumber  \\
\rho^2 \, g^{\psi \psi} = {}& \frac{ b^2 g^2 q^2}{ r^2 \Delta_r } +
\frac{\Xi_b}{\cos^2 \theta} + \frac{\Xi_b}{ r^2 \Delta_r } (1 + g^2
r^2) (r^2+ a^2) (a^2- b^2)   \nonumber \\
& - \frac{2 m}{ r^2 \Delta_r } (b^2 g^2 r^2
+ a^2) - \frac{2 a b q}{ \Xi_a \, r^2 \Delta_r } (\Xi_a \, g^2 (r^2 -
b^2) - a^4 g^4 +1)\,, \nonumber  \\
\rho^2 \, g^{\varphi \psi} = {}&  \frac{ a b g^2 q^2 - (1+ g^2 r^2) (2 m a b
  + (a^2 + b^2) q) }{ r^2 \Delta_r }\,, \nonumber \\
\rho^2 g^{\theta \theta} = {}& \Delta_{\theta}, \qquad \rho^2 g^{r r}
=\Delta_{r}\,. \label{inverse metric}
\end{align}
With \eqref{inverse metric}, transforming to the tortoise coordinate $r_*$ defined in \eqref{tortoise} and taking the near horizon limit ($\Delta_{r}\to 0$) the action in \eqref{free action with g's} becomes
\begin{align}
S&\approx\frac{1}{2\Xi_a\Xi_b}\int dt dr_* \sin2\theta d\theta d\varphi d\psi \frac{r^3}{(r^2+a^2)(r^2+b^2)+abq}\, \phi^* \Big(\tilde{g}^{t t}D_t^2+2\tilde{g}^{t \varphi}D_tD_\varphi+2\tilde{g}^{t \psi}D_tD_\psi \nonumber \\
&\quad\qquad\qquad +\tilde{g}^{\varphi \varphi}D_\varphi^2+\tilde{g}^{\psi \psi}D_\psi^2+2\tilde{g}^{\varphi \psi}D_\varphi D_\psi+\frac{\left[(r^2+a^2)(r^2+b^2)+abq\right]^2}{r^4}\p_{r_*}^2 \Big)\, \phi \,, \label{free near horizon action with gtildes}
\end{align}
where the $\tilde{g}$'s may be put into the following form:
\bea
r^4 \tilde{g}^{t t}&=&-\left[(r^2+a^2)(r^2+b^2)+abq\right]^2\,, \nonumber \\
r^4 \tilde{g}^{t \varphi}&=&-\left[(r^2+a^2)(r^2+b^2)+abq\right]\left[a(r^2+b^2)(1+g^2r^2)+qb\right]\,, \nonumber \\
r^4 \tilde{g}^{t \psi}&=&-\left[(r^2+a^2)(r^2+b^2)+abq\right]\left[b(r^2+a^2)(1+g^2r^2)+qa\right]\,, \nonumber \\
r^4 \tilde{g}^{\varphi \varphi}&=&-\left[a(r^2+b^2)(1+g^2r^2)+qb\right]^2\,, \nonumber \\
r^4 \tilde{g}^{\psi \psi}&=&-\left[b(r^2+a^2)(1+g^2r^2)+qa\right]^2\,, \nonumber \\
r^4 \tilde{g}^{\varphi \psi}&=&-\left[a(r^2+b^2)(1+g^2r^2)+qb\right]\left[b(r^2+a^2)(1+g^2r^2)+qa\right]\,.
\eea
The above clearly satisfy the following identities:
\bea
(\tilde{g}^{t \varphi})^2=\tilde{g}^{t t}\tilde{g}^{\varphi \varphi}\,, \nonumber \\
(\tilde{g}^{t \psi})^2=\tilde{g}^{t t}\tilde{g}^{\psi \psi}\,, \nonumber \\
\tilde{g}^{t \varphi}\tilde{g}^{t \psi}=\tilde{g}^{t t}\tilde{g}^{\varphi \psi}\,. \label{identities for gtildes}
\eea
The identities \eqref{identities for gtildes} along with the observation that $\Omega_a$ and $\Omega_b$ in \eqref{ang velocities} are given precisely by
\be
\Omega_{a} = \frac{\tilde{g}^{t \varphi}}{\tilde{g}^{t t}}\,, \qquad \Omega_{b} = \frac{\tilde{g}^{t \psi}}{\tilde{g}^{t t}}\,,
\ee
allow one to rewrite the action in \eqref{free near horizon action with gtildes} into the form given in the text \eqref{near horizon S with old theta}.
\bibliography{References}

\end{document}